\documentclass[showpacs]{revtex4}
\usepackage{epsfig}
\def\beq{\begin{equation}}
\def\enq{\end{equation}}
\def\beqa{\begin{eqnarray}}
\def\enqa{\end{eqnarray}}

\def\GeV{\nobreak\,\mbox{GeV}}

\def\qq{\lag\bar{q}q\rag}

\def\ss{\lag\bar{s}s\rag}
\def\mix{\lag\bar{q}g\si.Gq\rag}
\def\mixs{\lag\bar{s}g\si.Gs\rag}
\def\Gd{\lag g^2G^2\rag}
\def\G3{\lag g^3G^3\rag}

\def\rh{\rho}
\def\si{\sigma}

\def\al{\alpha}

\def\lb{\label}
\def\nn{\nonumber}

\def\qslash{\rlap{/}{q}}
\def\mxi{m_{\Xi_Q}}
\newcommand{\rag}{\rangle}
\newcommand{\lag}{\langle}

\begin{document}

\title{\sc
QCD sum rules study of $\Xi_c$ and $\Xi_b$ baryons
}
\author { Francisco O. Dur\~aes}
\email{duraes@mackenzie.com.br}
\affiliation{Centro de Ci\^encias e Humanidades,
Universidade Presbiteriana Mackenzie, R. da Consola\c{c}\~ao 
930, 01302-907 S\~ao Paulo, SP, Brazil}

\author{Marina Nielsen}
\email{mnielsen@if.usp.br}
\affiliation{Instituto de F\'{\i}sica, Universidade de S\~{a}o Paulo,
C.P. 66318, 05389-970 S\~{a}o Paulo, SP, Brazil}

\begin{abstract}
We use QCD  sum rules to study the masses of the baryons
$\Xi_c$ and $\Xi_b$. We work with a current where the strange and the light
quarks are in a relative spin zero,
at leading order in $\alpha_s$. We consider the contributions of
condensates up to dimension six.
For  $\Xi_b$ we get $m_{\Xi_{b}}= (5.75\pm 0.25)~{\rm GeV}$, 
and for  $\Xi_c$ we get $m_{\Xi_{c}}= (2.5\pm 0.2)~{\rm GeV}$, both 
in excelent agreement with the experimental values. We also make predictions
to the state $\Omega_b(ssb)$ obtaining $m_{\Omega_{b}}= (5.82\pm 0.23)~
{\rm GeV}$.
\end{abstract}

\maketitle

The recent observation of the $\Xi_b^-$ baryon by D0 and CDF collaborations
\cite{D0,litv} with mass in agreement with the prediction in ref.~\cite{kar},
has estimulated us to use the QCD sum rules (QCDSR) \cite{svz}
to evaluate the mass
of this state. Previous QCD sum rule calculations for beauty baryons have been
done before for the $\Lambda_b$, $\Sigma_b$, $\Sigma_b^*$ and for double
beauty baryons \cite{bcdn} but not for $\Xi_b^- (dsb)$. Since in the QCDSR 
approach, hadronic masses are related with the vaccum condensates, the use
of this method to evaluate hadronic masses is an important step in the
understanding of the dynamical nature of these masses.

Here we follow ref.~\cite{kar} and assume that the strange and light 
($sq$) quarks in $\Xi_b$ are in a relative spin zero state. Therefore,
the most general (low dimension) current which interpolates the $\Xi_b$ 
operator can be constructed from a combination between the two currents,
formed with scalar and pseudoscalar diquarks:
\beq
\eta_Q=\epsilon_{abc}\left[(q_a^TC\gamma_5s_b)+t(q_a^TCs_b)\gamma_5\right]Q_c,
\label{cur}
\enq
where $a,~b,~c$ are color indices, $C$ is the charge conjugation
matrix, $Q$ denotes the heavy quark and $t$ is the mixing parameter between 
the two currents. Of course the above interpolating 
field can also be used to study the $\Xi_c~(qsc)$ baryon.

The QCDSR is constructed from the two-point correlation function
\beq
\Pi(q)=i\int d^4x ~e^{iq.x}\lag 0
|T[\eta_Q(x)\bar{\eta}_Q(0)]|0\rag.
\lb{2po}
\enq
Lorentz covariance, parity and time reversal imply that the two-point
correlation function in Eq.~(\ref{2po}) has the form
\beq
\Pi(q)=\Pi_1(q^2)+q\kern-.5em\slash \Pi_2(q^2).
\lb{inv}
\enq
A sum rule for each invariant function $\Pi_1$ and $\Pi_2$, in Eq.(\ref{inv})
can be obtained.

The calculation of the phenomenological side at the 
hadron level proceeds by writing a dispersion relation to each one of the
invariant functions in Eq.~(\ref{inv}):
\beq
\Pi_i^{phen}(q^2)=-\int ds\, {\rho_i(s)\over q^2-s+i\epsilon}\,+\,\cdots\,,
\label{phen}
\enq
where $\rho_i$ is the spectral density and the dots represent subtraction 
terms. The spectral density is described, as usual, as a single sharp
pole representing the lowest resonance plus a smooth continuum representing
higher mass states:
\beqa
\rho_1(s)&=&\lambda^2m_{\Xi_Q}\delta(s-m_{\Xi_Q}^2) +\rho_1^{cont}(s)\,,\nn\\
\rho_2(s)&=&\lambda^2\delta(s-m_{\Xi_Q}^2) +\rho_2^{cont}(s)\,,
\label{den}
\enqa
where $\lambda^2$ gives the coupling of the current with
the low mass hadron of interest. For simplicity, it is
assumed that the continuum contribution to the spectral density,
$\rho_i^{cont}(s)$ in Eq.~(\ref{den}), vanishes bellow a certain continuum
threshold $s_0$. Above this threshold, it is assumed to be given by
the result obtained with the OPE. Therefore, one uses the ansatz \cite{io1}
\beq
\rho_i^{cont}(s)=\rho_i^{OPE}(s)\Theta(s-s_0)\;,
\enq
with
\beq
\rho_i^{OPE}(s)={1\over\pi}Im[\Pi_i^{OPE}(s)]\;.
\enq

On the OPE side, we work at leading order in $\alpha_s$ and consider the
contributions of condensates up to dimension six.  We keep the terms
which are linear in the strange-quark mass $m_s$.  We use the 
momentum space expression for the charm quark propagator,
while the light-quark part of the correlation function is
calculated in the coordinate-space. After
making a Borel transform of both sides, and
transferring the continuum contribution to the OPE side, the sum rules
for $\Xi_Q$ baryon, up to dimension-six condensates can
be written as:
\beq 
\lambda^2m_{\Xi_Q}e^{-{\mxi}^2/M^2}=\int_{m_Q^2}^{s_0}ds~
e^{-s/M^2}~\rho_1^{OPE}(s)\; +\Pi_1(M^2)\;, 
\lb{srm} 
\enq
\beq 
\lambda^2e^{-{\mxi}^2/M^2}=\int_{m_Q^2}^{s_0}ds~
e^{-s/M^2}~\rho_2^{OPE}(s)\; +\Pi_2(M^2)\;, 
\lb{srq} 
\enq
where
\beqa
\rho_i^{OPE}(s)&=&\rho_i^{pert}(s)+\rh_i^{\qq}(s)+\rh_i^{\lag G^2\rag}(s)\;,
\nn\\
\Pi_i(M^2)&=&\Pi_i^{\mix}(M^2)+\Pi_i^{\qq^2}(M^2).
\lb{rhoeq}
\enqa
In the structure 1 we get
\beqa\label{est1}
\rho_1^{pert}(s)&=&{(1-t^2)m_Q^5\over 2^{7} \pi^4}\left[(1-x)\left({1
\over x^2}+{10\over x}+1\right)+6\left(1+{1\over x}\right)\ln{x}\right],
\nn\\
\rho_1^{\qq}(s)&=&-{m_Qm_s\over2^3\pi^2}(1-x)\left[(1+t^2)\qq-{(1-t^2)\ss
\over2}\right],
\nn\\
\rho_1^{\lag G^2\rag}(s)&=&{(1-t^2)m_Q\Gd\over 2^{9}3 \pi^4}\left[(1-x)\left(7
+{2\over x}\right)+6\ln{x}\right],
\nn\\
\Pi_1^{\mix}(M^2)&=&{m_Qm_s\over2^5\pi^2}\left[{(1-t^2)\mixs\over6}
e^{-m_Q^2/M^2}-(1+t^2)\mix\left(e^{-m_Q^2/M^2}-\int_0^1d\al e^{-m_Q^2\over
(1-\al)M^2}\right)\right],
\nn\\
\Pi_1^{\qq^2}(M^2)&=&{m_Q\qq\ss\over6}(1+t^2)e^{-m_Q^2/M^2},
\enqa
where $x=m_Q^2/s$. In the structure $\qslash$ we get
\beqa\label{estq}
\rho_2^{pert}(s)&=&{(1+t^2)m_Q^4\over 2^{9}\pi^4}\left[(1-x^2)\left({1
\over x^2}-{8\over x}+1\right)-12\ln{x}\right],
\nn\\
\rho_2^{\qq}(s)&=&{m_s\over2^4\pi^2}(1-x^2)\left[-(1-t^2)\qq+{(1+t^2)\ss
\over2}\right],
\nn\\
\rho_2^{\lag G^2\rag}(s)&=&{(1+t^2)\Gd\over 2^{10}3 \pi^4}(1-x)(1+5x),
\nn\\
\Pi_2^{\mix}(M^2)&=&{m_s\over2^5\pi^2}\left[{(1+t^2)\mixs\over6}
e^{-m_Q^2/M^2}+(1-t^2)\mix\left(e^{-m_Q^2/M^2}+\int_0^1d\al(1-\al) 
e^{-m_Q^2\over(1-\al)M^2}\right)\right],
\nn\\
\Pi_2^{\qq^2}(M^2)&=&{\qq\ss\over6}(1-t^2)e^{-m_Q^2/M^2}.
\enqa

The contribution of dimension-six condensates $\lag g^3 G^3\rag$
is neglected, since it is assumed to be suppressed by   the loop
factor $1/16\pi^2$.

In the numerical analysis of the sum rules, the values used for the
quark
masses and condensates are (see e.g.  \cite{svz,bcdn}):
$m_s=(0.10\pm 0.03)\,\GeV$, $m_c(m_c)=(1.23\pm 0.05)\,\GeV $, 
$m_b(m_b)=(4.24\pm 0.06)\,\GeV$,
$\lag\bar{q}q\rag=\,-(0.23\pm0.03)^3\,\GeV^3$, $\ss=0.8\qq$,
$\lag\bar{q}g\si.Gq\rag=m_0^2\lag\bar{q}q\rag$ with $m_0^2=0.8\,\GeV^2$,
$\lag g^2G^2\rag=0.88~\GeV^4$.

\begin{figure}[h]
\centerline{\epsfig{figure=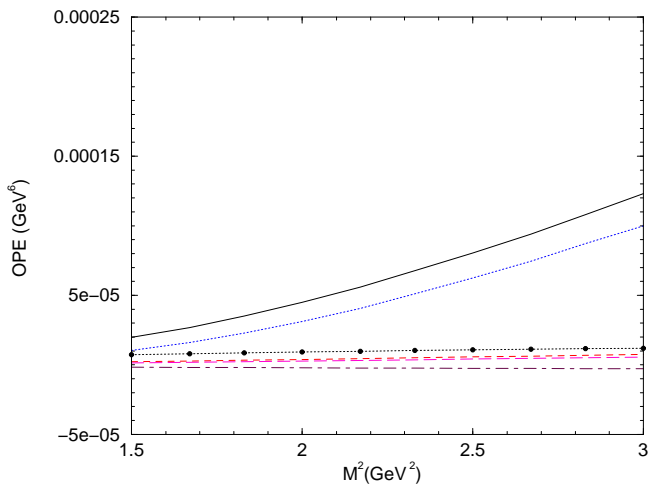,height=65mm}}
\caption{Same as Fig.~1 for $t=0$. The solid line with dots gives the 
four-quark condensate contribution}
\label{figconvc0}
\end{figure}

We start with the  charmed baryon $\Xi_c$. We evaluate the sum rules in the 
range $1.5\leq M^2 \leq 3.0~\GeV^2$ for
$s_0$ in the range: $3.0\leq \sqrt{s_0} \leq3.2$ GeV.
\begin{figure}[h]
\centerline{\epsfig{figure=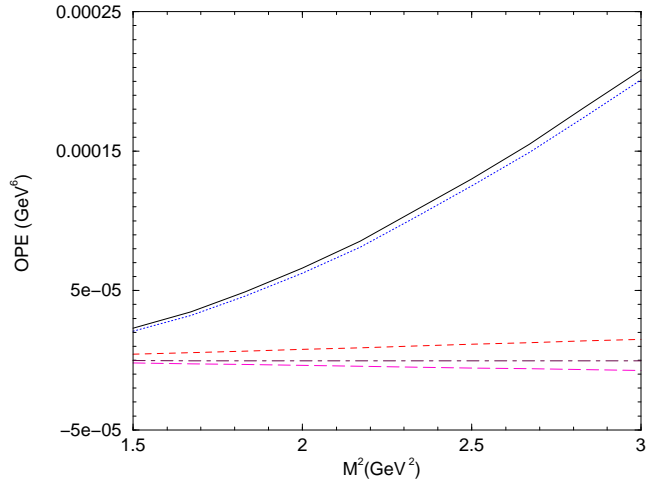,height=65mm}}
\caption{The OPE convergence for the sum rule Eq.(\ref{srq}) for $\Xi_c$,
using $\sqrt{s_0} = 3.1$ GeV 
and $t=1$. The dotted, long-dashed, dashed and dot-dashed lines give, 
respectively, the perturbative, quark condensate, gluon condensate and mixed 
condensate contributions. The solid line gives the total OPE contribution
to the sum rule.}
\label{figconvc1}
\end{figure}
In Fig.~\ref{figconvc1} we show 
the contribution of each term in Eq.~(\ref{estq}) to the sum rule in
Eq.~(\ref{srq}), for $t=1$ and $\sqrt{s_0}=3.1~\GeV$. We see that we get an 
excelent OPE convergence. For $t=1$ the four-quark condensate contribution
to the sum rule Eq.~(\ref{srq}) vanishes. 
For other values of $t$, although the four-quark condensate contribution 
is bigger than the contribution of the other 
condensates, it is still much smaller than the perturbative contribution and,
therefore, it does not spoil the convergence of the sum rule, as can be seen 
in Fig.~\ref{figconvc0}. 
Therefore we conclude that the convergence of the
sum rule in the $\qslash$ structure in Eq.~(\ref{srq}), is good for any 
value of the mixing parameter $t$ (see Eq.(\ref{cur})) in the range 
$0\leq t\leq1$. This is not the case of the sum rule in Eq.~(\ref{srm})
(structure 1), since for $t=1$ the perturbative and gluon condensate 
contributions vanishe. As a matter of fact, if we try to obtain the mass
of the $\Xi_c$ baryon by dividing Eq.~(\ref{srm}) by Eq.~(\ref{srq}), we
only get values compatible it the experimental mass for $t\sim0$. Therefore,
in this work we will use only the sum rule in Eq.~(\ref{srq}). To obtain the
mass of the baryon we take the derivative of Eq.~(\ref{srq})
with respect to $1/M^2$, divide the result by Eq.~(\ref{srq}) and
obtain:
\beq
m_{\Xi_Q}^2={\int_{m_Q^2}^{s_0}ds ~e^{-s/M^2}~s~\rho_2^{OPE}(s)-(d\Pi_2
/dM^{-2})
\over\int_{m_Q^2}^{s_0}ds ~e^{-s/M^2}~\rho_2^{OPE}(s)+\Pi_2(M^2)}\;.
\lb{m2}
\enq

\begin{figure}[h]
\centerline{\epsfig{figure=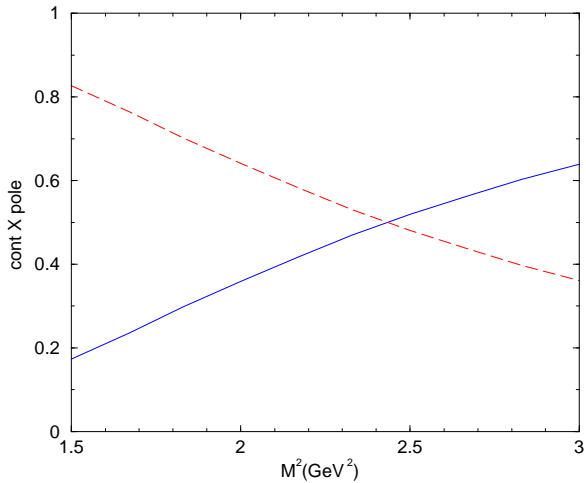,height=65mm}}
\caption{The dashed line shows the relative pole contribution (the
pole contribution divided by the total, pole plus continuum,
contribution) and the solid line shows the relative continuum
contribution for $\sqrt{s_0}=3.1~\GeV$ and $t=1$.}
\label{figpvc}
\end{figure}

We get an upper limit constraint for $M^2$ by imposing the rigorous
constraint that the QCD continuum contribution should be smaller than the
pole contribution.
The maximum value of $M^2$ for which this constraint is satisfied
depends on the value of $s_0$ and $t$.  The comparison between pole and
continuum contributions for $\sqrt{s_0} = 3.1$ GeV  and $t=1$ is shown in
Fig.~\ref{figpvc}. 
The same analysis for the other values of the continuum threshold and $t=1$
gives $M^2 \leq 2.3$  GeV$^2$ for $\sqrt{s_0} = 3.0~\GeV$ and
$M^2 \leq 2.65$  GeV$^2$ for $\sqrt{s_0} = 3.2~\GeV$. We get similar results 
for other values of $t$, for example for $\sqrt{s_0} = 3.1~\GeV$ and
$t=0$ we get $M^2 \leq 2.6~\GeV^2$.

\begin{figure}[h]
\centerline{\epsfig{figure=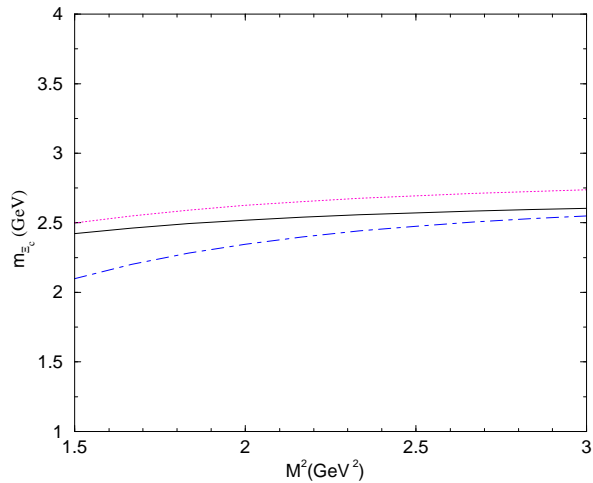,height=65mm}}
\caption{The $\Xi_c$ baryon mass as a function of the Borel parameter
($M^2$) for different values of the continuum threshold and the current mixing
parameter: $t=1$, $\sqrt{s_0} =3.2$ GeV (solid line); $t=1$, $\sqrt{s_0} = 
3.0$ GeV (dotted line); $t=0$, $\sqrt{s_0} =3.1$ GeV (dot-dashed line).}  
\label{figmx}
\end{figure}

In Fig.~\ref{figmx}, we show the $\Xi_c$ baryon mass obtained from
Eq.~(\ref{m2}), in the relevant sum rules window for different values
of $\sqrt{s_0}$ and $t$. From Fig.~\ref{figmx} we see that the results are 
more  stable, as a function of $M^2$, for $t=1$ than for $t=0$. Therefore,
we will use $t=1$ to estimate the mass of the particle. It is very interesting
to notice that the result for the mass obtained with $t=0$ is very similar 
to the one obtained dividing Eq.~(\ref{srm}) by Eq.~(\ref{srq}) using
also $t=0$.

We found that our results are not very sensitive to the value of the
charm quark mass, neither to the value of the condensates. The most
important source of uncertainty is the value
of the continuum threshod and the Borel interval. Using the QCD parameters
given above, the QCDSR result for the $\Xi_c$ baryon mass is:
\beq
m_{\Xi_c} = (2.5\pm0.2)~\GeV,
\enq
in a very good agreement with the experimental value $m_{\Xi_c}^{exp}=
(2.4710\pm0.0004)~\GeV$ \cite{pdg}.


In the case of the beauty baryon $\Xi_{b}$,
using consistently the perturbative $\overline{MS}$-mass $m_b(m_b)=(4.24
\pm0.6)~\GeV$, $t=1$
and the continuum threshold in the range $6.3\leq\sqrt{s_0}\leq6.5~\GeV$,
we find a good OPE convergence for $M^2>4.0~\GeV^2$.
We also find that the pole contribution is bigger than the continuum
contribution for $M^2<5.2~\GeV^2$ for $\sqrt{s_0}<6.3~\GeV$, and
for $M^2<5.7~\GeV^2$ for $\sqrt{s_0}=6.5~\GeV$.

\begin{figure}[h]
\centerline{\epsfig{figure=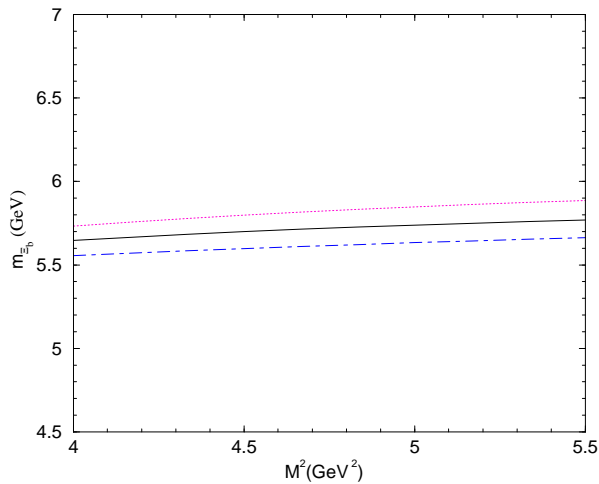,height=65mm}}
\caption{The $\Xi_b$ baryon mass as a function of the Borel parameter
($M^2$) for different values of the continuum threshold and the current mixing
parameter: $t=1$, $\sqrt{s_0} =6.3$ GeV (solid line); $t=1$, $\sqrt{s_0} = 
6.5$ GeV (dotted line); $t=0$, $\sqrt{s_0} =6.4$ GeV (dot-dashed line).}
\label{figmxb}
\end{figure}

From Fig.~\ref{figmxb}, where  we show the $\Xi_{b}$ baryon mass, 
we see that the results are very stable as a
function of $M^2$ in the allowed Borel region. For completeness we also show,
in Fig.~\ref{figmxb}, the results obtained using $t=0$ and $\sqrt{s_0} =6.4$ 
GeV. In this case the results are very stable for all values of 
$t$ in the range $0\leq t\leq1$. Therefore, we will also use different values
of  $t$ to estimate the uncertainties in the result.
Taking into account the
variation of $M^2$ and varying $s_0$, $t$ and $m_b$ in the regions indicated
above, we arrive at the result:
\beq
\lb{massXb}
m_{\Xi_{b}}=  (5.75\pm0.25)~\GeV~,
\enq
also in a very good agreement with the predictions in refs.~\cite{kar}
and \cite{jen}, and with the experimental results in ref.~\cite{D0}:
$m_{\Xi_b}^{D0}=(5.774\pm0.013)~\GeV$, 
and in ref.~\cite{litv}: $m_{\Xi_b}^{CDF}=(5.7929\pm0.0024)~\GeV$ 

We have presented a QCDSR analysis of the two-point
functions of the $\Xi_Q(qsQ)$ baryons.  We find that the sum rules results 
for the masses of
$\Xi_{c}$ and $\Xi_{b}$ are compatible with the experimental values and
with the predicitons in refs.~\cite{kar,jen}, in the case of $\Xi_{b}$.
These results for $\Xi_{b}$ are sumarized in Table I.

\begin{center}
\small{{\bf Table I:} Results for the $\Xi_b$ mass.}
\\
\begin{tabular}{|c|c|}  \hline
$m_{\Xi_b}~(\GeV)$ & ref.  \\
\hline
 $5.774\pm.011\pm0.015$ & D0 \cite{D0} \\
\hline
$ 5.7929\pm0.0024\pm0.0017$ & CDF \cite{litv} \\
\hline
$ 5.795\pm0.005$ &\cite{kar}\\
\hline
$ 5.8057\pm0.0081$ &\cite{jen}\\
\hline
$ 5.75\pm0.25$ &this work\\
\hline
\end{tabular}\end{center}

It is important to notice that while the calculation based on modeling
hyperfine interaction \cite{kar,jen} needs, as inputs, the masses of others 
heavy baryons, in our calculations the masses are extract using only  
information about the QCD parameters as quark masses and condensates. In
the case of the heavy baryons $\Xi_{c}$ and $\Xi_{b}$, in particular, their
masses are determined basicaly by the first term in the sum rules Eqs.
(\ref{srm}) and (\ref{srq}) and the condensates are not very important.

We have tested two different choices of currents. While the
results for $\Xi_c$ are sensitive to this choice, in the case of
$\Xi_b$ the results are very stable for the mixing parameter in the range
$0\leq t\leq1$.

It is not possible to generalize directly our results to the baryons 
$\Omega_c(ssc)$ and $\Omega_b(ssb)$, from the current in Eq.(\ref{cur}), 
since one can not construct an
scalar or pseudoscalar diquark in a $\bar{3}$ configuration of color, with two 
strange quarks. Therefore, to study the baryon $\Omega_Q$ we can use a 
proton-like current:
\beq
\eta_\Omega=\epsilon_{abc}(s_a^TC\gamma_\mu s_b)\gamma_5\gamma^\mu Q_c.
\label{cur2}
\enq
With this current, the OPE contributions to the sum rule in the structure 
$\qslash$ (Eq.~(\ref{srq})) for the $\Omega_Q$ baryon, up to dimension-six 
condensates, are given by:
\beqa\label{omega}
\rho_2^{pert}(s)&=&{m_Q^4\over 2^{6}\pi^4}\left[(1-x^2)\left({1
\over x^2}-{8\over x}+1\right)-12\ln{x}\right],
\nn\\
\rho_2^{\qq}(s)&=&0,
\nn\\
\Pi_2^{\lag G^2\rag}(M^2)&=&-{m_Q^2\Gd\over 2^{16}3 \pi^4},\int_0^1d\al
{\al^2\over(1-\al)^2} e^{-m_Q^2/M^2},
\nn\\
\Pi_2^{\mix}(M^2)&=&-{7m_s\over48\pi^2}\mixs e^{-m_Q^2/M^2},
\nn\\
\Pi_2^{\qq^2}(M^2)&=&{2\ss^2\over3}e^{-m_Q^2/M^2}.
\enqa

We find that the OPE convergence of this sum rule is as good as the OPE 
convergence obtained for $\Xi_Q$ in the same Borel region: $M^2\geq 1.5
~\GeV^2$ for $\Omega_c$ and $M^2\geq4.0~\GeV^2$ for $\Omega_b$. In  
Table II we give the maximum value of $M^2$ for which the continuum 
contribution is smaller than 50\%, for different values of the continuum
threshold.

\begin{center}
\small{{\bf Table II:} Maximum values of the Borel parameter $(M^2$)
for different values of  $s_0$.}
\\
\begin{tabular}{|c|c|c|}  \hline
state & $\sqrt{s_0}(\GeV)$ &$M^2_{max}(\GeV^2)$  \\
\hline
 $\Omega_c$ & 3.2 & 2.7 \\
\hline
 $\Omega_c$ & 3.4 & 3.1 \\
\hline
 $\Omega_b$ & 6.4 & 5.5 \\
\hline
 $\Omega_b$ & 6.7 & 6.5 \\
\hline
\end{tabular}\end{center}

The results obtained for $m_{\Omega_Q}$, from Eq.(\ref{m2}), are also very 
stable, as a function of $M^2$, in the allowed Borel window. Considering
the variations in $s_0$ and $M^2$ given in Table II, and the variations in the
quark masses and condensates, as discussed above, we get
\beq
m_{\Omega_{c}}=  (2.65\pm0.25)~\GeV~,
\enq
in a very good agreement with the experimental value $m_{\Omega_{c}}^{exp}=
(2.6975\pm0.0026)~\GeV$ \cite{pdg}. For $\Omega_b$ we make the prediction:
\beq
m_{\Omega_{b}}=  (5.82\pm0.23)~\GeV~.
\enq

As a final remark, it is very reasurring to see that the OPE convergence is so 
good for heavy baryons, since this is not the case for tetraquark states
\cite{tetra}. As shown in ref.~\cite{tetra}, it is very difficult to find a 
Borel region where the continuum contribution is bigger than the pole 
contribution and where the OPE convergence is acceptable, for tetraquark 
states with only one heavy quark. Therefore, from a QCD sum rule point of 
view, it is much easier
to form an state, separated from the continuum,  with three quarks than 
with  four quarks.

\section*{Acknowledgements}
{This work has been partly supported by FAPESP and CNPq-Brazil.}


\end{document}